\documentclass[12pt]{article}

\def\rdots{\mathinner{\mkern1mu\raise1pt\vbox{\kern1pt\hbox{.}}\mkern2mu
   \raise4pt\hbox{.}\mkern2mu\raise7pt\hbox{.}\mkern1mu}}
\newcommand{\Z}{{\rm Z\kern-.35em Z}}
\newcommand{\bP}{{\rm I\kern-.15em P}}
\newcommand{\Q}{\kern.3em\rule{.07em}{.65em}\kern-.3em{\rm Q}}
\newcommand{\R}{{\rm I\kern-.15em R}}
\newcommand{\h}{{\rm I\kern-.15em H}}
\newcommand{\C}{\kern.3em\rule{.07em}{.65em}\kern-.3em{\rm C}}
\newcommand{\T}{{\rm T\kern-.35em T}}

\newcommand{\be}{\begin{equation}}
\newcommand{\ee}{\end{equation}}

\newcommand{\pa}{\partial}
\newcommand{\cd}{\cdot}
\newcommand{\ra}{\rightarrow}

\newcommand{\de}{\delta}
\newcommand{\Del}{\Delta}
\newcommand{\La}{\Lambda}

\newtheorem{theorem}{Theorem}

\begin{document}
%\font\twelverm=cmr12
%\font\cs=CMSSI12

\openup 1.5\jot
\centerline{A random walk on the permutation group, some formal long-time asymptotic relations}

\vspace{1in}
\centerline{Paul Federbush}
\centerline{Department of Mathematics}
\centerline{University of Michigan}
\centerline{Ann Arbor, MI 48109-1109}
\centerline{(pfed@umich.edu)}

\vspace{1in}

\centerline{\underline{Abstract}}

We consider the group of permutations of the vertices of a lattice.  A random walk is generated by unit steps that each interchange two nearest neighbor vertices of the lattice.  We study the heat equation on the permutation group, using the Laplacian associated to the random walk.  At $t=0$ we take as initial conditions a probability distribution concentrated at the identity.  A natural conjecture for the probability distribution at long times is that it is 'approximately' a product of Gaussian distributions for each vertex.  That is, each vertex diffuses independently of the others.  We obtain some formal asymptotic results in this direction.  The problem arises in certain ways of treating the Heisenberg model in statistical mechanics.

\vfill\eject

This paper represents work in progress, and is written in a casual manner, as notes for a seminar or perhaps a physics article.  But the work is interesting and non-trivial, and perhaps will inspire research, many natural questions will appear.  We of course will be clear about what is proved and what is conjectured, some proofs will be sketched.  As will be seen this work is somewhat discouraging about the application sought to the Heinsenberg model problem, but opens some pleasant mathematical vistas.

We study a d-dimensional periodic lattice cube, $\Lambda$ of edge size $L$.  The number of its vertices, $\cal V$, is given as
\be	|\Lambda| \equiv N \equiv L^d \equiv \# \{ {\cal V} \}.	\ee
We set ${\cal G}$ to be the permutation group on $\cal V$.  $H$ is the element in the group algebra given as
\be	H = - \sum_{i \sim j } (I_{ij} - I).	\ee
here $i\sim j$ indicates that $i$ and $j$ are neighboring vertices in the lattice. $I$ is the identity element in $\cal G$, and $I_{ij}$ is the group element that interchanges vertices $i$ and $j$ leaving the other vertices alone.  $H$ is thus the Heisenberg model Hamiltonian ``promoted" from an operator in the Hilbert space to an element of the group algebra.  (One need not be familiar with the Heisenberg model for purposes of this paper.)

We then consider the group algebra element $e^{-Ht}$.  This can be expanded as a linear combination of group elements, the $g_p$
\be     e^{-Ht} = \sum_p \ f(g_p, t)g_p \ee
where $p$ labels the group elements.  Each $g_p$ represents a mapping of the vertices,
\be		g_p : i \rightarrow p(i)	\ee
so $p$ is specified by
\be   \Big( p(1), p(2), ..... \Big)	\ee
Equation (5) is a point in $(Z^d)^N$, in fact in $(\Lambda)^N$.  The cardinality of the set of such points is $N!$.  Such points in $(\Lambda)^N$ are restricted by the condition that all the $p(i)$ in (5) are distinct.  This subset we label $(\Lambda)^{N*}$.  The respective cardinalities of $(\Lambda)^{N*}$ and $(\Lambda)^{N}$ are $N!$ and $N^N$.  The latter space is simpler, being a periodic lattice cube.

We have a time dependent probability function $f(g,t)$ on the permutation group (or on $(\Lambda)^{N*}$).  We now extend $f(g,t)$ to $f^e(\vec{x}, t), \vec{x}$ in $(\Lambda)^{N}$.  Of course such extensions are not unique.  The motivation for extending $f$ will soon be clear, largely depending on the simplicity of $(\Lambda)^{N}$ over $(\Lambda)^{N*}$.  The extended function will no longer be a probability function.

The relation between the Heisenberg model of ferromagnetism and our random walk on the permutation group was beautifully developed by R. Powers in [1].  Inspired by this work, the author presented a possible avenue towards proving the phase transition of the Heisenberg model in [2], a development of ideas in [1].  The central relation needed in the proof envisioned in [2] is of the form
\be    f^e(\vec{x}, t) \cong C_N \prod_{i\in {\cal V}} \ (e^{\Delta t})_{i,x_i} \ \ {\rm for} \ t \ {\rm large}.     \ee
$\Delta$ is the lattice Laplacian on $\Lambda$.  (Of course, equation (6) need only hold on $f(g,t)$, but we presume the extension satisfies (6).)  We do not now make explicit the degree of approximation implied by $\cong$.  The right side of (6) is a product of gaussians (associated to independent random walks).  It also is a solution of the heat equation on $(\Lambda)^{N}$ with its natural lattice Laplacian!

At the very least we would want (6) to imply
\be \lim_{t \ra \infty}  f^e(\vec{x}, t) = C_N \lim_{t \ra \infty} \prod_{i\in {\cal V}} \ (e^{\Delta t})_{i,x_i}. \ee
We note
\be   \lim_{t \ra \infty} f(g,t) = \frac 1{N!}    \ee
and
\be   \lim_{t \ra \infty} (e^{\Delta t})_{i,j} = \frac 1{N}.   \ee
Restricting to points $\vec x$ in $(\La)^{N*}$ equation (7) becomes
\be	\frac 1{N!} = C_N\left(\frac 1 N\right)^N	\ee
From (10) and Stirling's formula we have determined the $C_N$ in (6) satisfy
\be 	\lim_{N \ra \infty} (C_N)^{\frac 1 N} = e	\ee

We will find an extension $f^e$ of $f$ satisfying a differential equation
\be		\frac {\pa f^e}{\pa t} = \Delta \ f^e + V \ f^e	\ee
$\Delta$ is the usual lattice Laplacian on $(\La)^N$ a periodic version of the lattice Laplacian on $Z^{dN}$.  Here we will only consider $V$ built up with ``two-particle" interactions.  Thus
\be	V = \sum \ V_{i,j}	\ee
where $V_{i,j}$ describes the interaction of the two vertices that at time $t=0$ were at $i$ and $j$ respectively.  The $V$ must be such that the solution of (12) (with initial conditions, the identity at $t=0$) restricted to $(\La)^{N*}$ agrees with $f$ as defined in equation (3).  It indeed is possible to find such $V$ that achieve this.

A form for $V_{i,j}$ that works is given as follows.  We apply this $V$ to a product function
\be	\phi(\vec x) = \prod_i \ \phi_i(x_i)	\ee
($V$ applied to product functions determines $V$ uniquely.)
\[	\left(V_{i,j} \phi \right) (\vec x) =  - \; \prod_{k\not= i,j} \phi_k(x_k) \cdot \sum_{y \in {\cal V}} \cdot \sum_{{\vec i}} \cdot \]
\[	\cdot \left[ \de_{x_i,y} \; \de_{x_j,y+{\vec i}} + \de_{x_j,y} \de_{x_i,y+{\vec i}} + r\; \de_{x_i,y}\de_{x_j,y} + r \;
\de_{x_i,y+{\vec i}} \de_{x_j,y+{\vec i}} \right] \cdot \]
\be \cdot \left[\phi_i(y) - \phi_i(y+{\vec i})\right] \cd \left[ \phi_j(y) - \phi_j(y + \vec i)\right]  \ee
$r$ is arbitrary.  There are $d$ orthonormal unit vectors, $\vec i$.  So the paths interact only when the vertices are in nearest neighbor position.  The expression (15) certainly is uniquely determined by the conditions above.  This is straightforward to show....though I labored weeks on it.  Herein we work with $r$ set equal to zero.  (But we believe working with an $r \not = 0$ at least if $|r| < 1$, leads to no essential changes in the form of our calculations and results.  We have studied this $r\not= 0$ situation a little, one need consider more diagrams than in the $r=0$ case.  We will briefly comment on this again later.)

We return to equation (6) for some further deliberations.  We consider summing $\vec {x}$ over $(\La)^N \equiv A$ and over $(\La)^{N*} \equiv B$.
\be
\sum_{\vec {x} \in A} f^e(\vec{x}, t) \cong C_N \sum_{\vec {x} \in A} \ \prod_{i\in {\cal V}}  (e^{\Delta t})_{i,x_i} = C_N  \ {\rm for} \ t \ {\rm large}
\ee
\be
\sum_{\vec {x} \in B} f^e(\vec{x}, t) = \sum_{\vec {x} \in B} f(\vec{x}, t) = 1 \cong C_N \sum_{\vec {x} \in B} \ \prod_{i\in {\cal V}}  (e^{\Delta t})_{i,x_i}  \ {\rm for} \ t \ {\rm large}
\ee
We will want the content of (16) and (17) to be given by the following conjectures.
\bigskip
\centerline{--------------------}
\underline{Conjecture 1}
\be   \lim_{t\ra \infty} \left( \sum_{\vec{x} \in A} f^e(\vec{x}, t)\right)^{1/N} = \left(C_N\right)^{1/N}  \ee
where $C_N$ is given by (10) and the limit is uniform in $N$.

\bigskip
\centerline{--------------------}

It is the uniformity requirement that makes the conjecture most difficult.
\bigskip
\centerline{--------------------}

\underline{Conjecture 2}

\be   \lim_{t\ra \infty} \left( \sum_{\vec{x} \in B} \ \prod_{i\in {\cal V}}  (e^{\Delta t})_{i,x_i}\right)^{1/N} = \left(C_N\right)^{-\; \frac1 N}  \ee
where $C_N$ is given by (10) and the limit is uniform in $N$.

\centerline{--------------------}

We believe Conjecture 2 is not very difficult to prove, and we plan to turn to it soon.

We now consider the solution of (12), treating $V$ as a perturbation in the form of a Rayleigh-Schrodinger expansion.
\be
f^e(\vec x, t) = \sum^\infty_{n=0} \int^t_0 dt_n \int^{t_n}_0 dt_{n-1} \cdots \int^{t_2}_0 dt_1 \ e^{\Del(t-t_n)} Ve^{\Del(t_n-t_{n-1})} \cdots Ve^{\Del t_1}.	\ee
The right side of (20) is naturally represented as a sum of contributions of diagrams.  In some more detail as a sum of products of the contributions of connected diagrams.  All the computations of this paper deal with results for the sum over final states (in (20) over $\vec x$ in $(\La)^N)$ for connected tree-graph diagrams.  If in equation (6) we sum over final states, over $\vec x$, on both sides we get
\be	\sum_{{\vec x}} f^e(\vec x, t) \cong C_N \ \ \ {\rm for} \ \ t \ \ {\rm large}.	\ee
(See Conjecture 1.)

Yes, in this paper we study the truth of (21), a much weakened form of (6).  But we expect that if we can get on top of (21) we are well on our way to treating (6).  Studying the decomposition of (20) into connected diagram contributions puts us in the ``cluster expansion" framework familiar in statistical mechanics.  A very complete treatment is in [3], but the level of sophistication of an undergraduate course in statistical mechanics is more than adequate.

\bigskip

\centerline{--------------------}

\begin{theorem}
We consider a two-particle (vertex) connected tree-graph contribution to (20).  The diagram is an $H$ shaped figure.  The bottom legs end at vertices $i$ and $j$ and the upper arms end at $x_i$ and ${x_j}$.  The cross segment represents an interaction at time $t_1$.  The contribution of this diagram, with $x_i$ and $x_j$ summed over ${\La^2}$, but before integrating over $t_1$ is
\be \sum_y \frac d{dt} \left( \phi_1(y,t) \ \phi_2(y,t)\right)\bigg|_{t=t_1}   \ee
where $\phi_1$ satisfies
\begin{eqnarray}
\frac \pa{\pa t} \phi_1 &=& \Del \phi_1 \\
\phi_1(x,0) &=& \de_{x,i} 
\end{eqnarray}
with $\Del$ in (23) the Laplacian in $\La$.  Similarly for $\phi_2$.
\end{theorem}

\centerline{--------------------}

This is a simple computation patterned on the continuum or lattice version of
\be
-2 \int \vec \nabla \phi_1 \vec \nabla \phi_2 = \int \left( \phi_1 \Del \phi_2 + \phi_2 \Del \phi_1 \right) = 
\int \left( \dot \phi_1 \phi_2 + \phi_2 \dot \phi_1 \right) = \int \frac d{dt} (\phi_1 \phi_2)
\ee

We will always be dealing with tree-graph diagrams.  Non tree-graph diagrams have contributions that fall off with $t$ (for $r = 0)$, and we are considering the $t \ra \infty$ limit.  We do not control the uniformity of this limit for the sum over all such diagrams, one reason for the formal nature of our computations.  Our final comment on the $r \not= 0$ case is that corresponding to Theorem 1, in this case, one must consider all ``ladder" diagrams to get the same formal estimate, and not just the single tree graph.  Choosing $r \not= 0$ leads to much more work and no gain.  To do better than our "results", if that is possible, one must consider potentials $V$ with other than two-body forces.
\bigskip
\centerline{-----------------------}

Theorem 2 will be the analog of Theorem 1 for connected diagrams involving $n$ particles, $n$ not necessarily 2.

\bigskip
\centerline{-----------------------}

\begin{theorem}
We consider all $n$-particle (vertex) connected tree-graph contributions to (20) involving vertices $z_1,z_2,...,z_n$ at $t=0$.  We sum over final positions, over $\La^n$.  We do not integrate over $t_1$, and over the other times in the order $t_n, t_{n-1}, ..., t_2$. Using Theorem 1, and Theorem 2 inductively on $n$, each of these integrals will be of an explicit time derivative.  In evaluating these integrals \underline{we keep only the lower limit}, as if in
\be	\int^b_a f'(t)dt = f(b) - f(a)	\ee
we keep only the $-f(a)$ term.  We will collect contributions of the upper limits later.  The ``contribution of lower limits" is
\be  (-1)^N (n-1)! \frac d{dt} \sum_y \ \prod^n_{i=1} \ \phi_i(y) \bigg|_{t=t_1}	\ee
where
\begin{eqnarray}
\frac \pa{\pa t} \; \phi_i &=& \Delta \phi_i	\\
\phi_i(y,0) &=& \delta_{y,z_i}
\end{eqnarray}
\end{theorem}
\bigskip
\centerline{---------------------}
\bigskip
\underline{Proof.}  The proof is by induction on $n$.  Let the earliest interaction corresponding to $t_1$ be between vertices $n-1$ and $n$.  (That is, one of these vertices at $z_n$ at $t=0$ and the other at $z_{n-1}$.)  Upon integrating over the later times and keeping only lower limits we have terms with vertices $1,2,...,j$ at $y$, and vertices $j+1, ..., n-2$ at $y + \vec i$, where vertices $n$ and $n-1$ are at these two points at $t_1$.  One sums over the value of $j$, the points $y$, the unit vectors $\vec i$, and the permutation of different possibilities for the vertices attached to $y$ and $y+\vec i$.  We let $S$ stand for the sum of the $(n-1)!$ permutations of vertices $1,...,n-1$.  The following telescopic sum relation is the heart of the proof.
\[	\sum_y \sum_{\vec i} S \left\{ \left[ \sum^{n-2}_{j=0} \phi_1(y) .... \phi_j(y) \phi_{j+1}(y+\vec i) ... \phi_{n-2}(y+ \vec i)\right] \cd \left(\phi_{n-1}(y) - \phi_{n-1}(y+\vec i)\right) \right\} \left(\phi_n(y) - \phi_n(y+\vec i)\right)  \]
\[  = \sum_y \sum_{\vec i} S \left[ \phi_1(y) ... \phi_{n-1}(y) - \phi_1(y+\vec i) ... \phi_{n-1}(y + \vec i)\right]  
 \left(\phi_n(y) - \phi_n(y+\vec i)\right)  \]
\be = \sum_y \sum_{\vec i} (n-1)! \left[ \phi_1(y) ... \phi_{n-1}(y) - \phi_1(y+\vec i) ... \phi_{n-1}(y + \vec i)\right]  
 \left(\phi_n(y) - \phi_n(y+\vec i)\right) 
\ee
This will represent the contribution from terms where the first interaction involves vertex $n$.  One then sums over the $n$ possibilities for this first vertex coupled.  (There is a factor of 2 in the first, $t=t_1$, interaction we have absorbed against the fact that we are counting double since either end of the interaction at $t=t_1$, could have been called $n$.)  Thus the proof is short.  Even writing out the details which we have raced over.  But the proof is tricky enough, so that it's hard to be sure you're right.  Counting is hard.

\centerline{--------------}

Theorem 1 may be included in the statement of Theorem 2 as the $n=2$ case.  In the next theorem we take the same contributions as in Theorem 2 but in addition integrate over $t_1$ from $0$ to $t$, and sum over $z_2,...,z_n$ but requiring that $z_1,...,z_n$ be a point in $(\La)^{n*}$.  We first define
\be   a_i \equiv - (-1)^i.	\ee

\centerline{--------------}

\begin{theorem} Let $T_n(t)$ be all the contributions considered in Theorem 2 for given $n$ in addition integrated over $t_1$ from 0 to $t$, and summed over $z_2,...,z_n$ with the requirement that $z_1,...,z_n$ be a point in $(\La)^{n*}$.
\be	\lim_{t\ra \infty} T_n(t) = a_{n-1}	\ee
\end{theorem}

\centerline{--------------}

\underline{Proof.}  We detail the proof for $n=2$ which contains all the essential points.  From (27) to (29) we have upon integrating (27) from 0 to $t$ and summing over $z_2$
\be \sum_{z_2 \not= z_1} \left( \sum_y \ \phi_1(y,t) \phi_2(y,t) - \sum_y \ \phi_1(0,t) \phi_2(0,t) \right)  \ee
with $\phi_1(y,0) = \de_{y,z_1}, \ \ \phi_2(y,0) = \de_{y,z_2}$.

\noindent
The second term in (33) therefore vanishes, leaving
\be	 \sum_{z_2 \not= z_1}  \sum_y \phi_1(y,t) \phi_2(y,t)  \ee
which equals
\be	 \sum_{z_2}  \sum_y \phi_1(y,t) \phi_2(y,t)  - \sum_y \ \phi^2_1(y,t)  .  \ee
The first term is 1 by
\begin{eqnarray}
 \sum_{z_2}   \phi_2(y,t) &=& 1 \\
\sum_y   \phi_1(y,t)  &=& 1.
\end{eqnarray}
and the second term in (35) gets to zero with $t$.  This concludes the proof for $n=2$.  For $n=3$, say, the initial points $z_1, z_2, z_3$ and $z_1, z_3, z_2$ lead to the same set of diagram contributions explaining the lost $(n-1)!$ factors when pursued.

We now define quantities $A_i$ defined recursively from the $a_i$ of (31).  We set
\be P \equiv \sum_{j=0} \ A_j \ t^j  \ee
a formal power series in $t$.  Then the $A_i$
are defined by
\be	A_0 = 1 \ee
\be A_i = a_i + \sum^{i-1}_{k=1} a_k \ {\rm coef} \ \left(P^{2k+1}, t^{i-k} \right), \ i > 0    \ee
where using Maple notation, coef $(f,t^s)$ picks out the coefficient of $t^s$ in the formal power series $f$.  This is the procedure by which we found the $A_i$.  Actually, with $A_1 = 1, A_2 = 2, A_3 =5, A_4 =14....$, the $A_i$ are the Catalan numbers, given as
\be	A_i = \frac 1{i+1} \left( \begin{array}{c}
2i \\
 \ \\
i
\end{array} \right). \ee

\centerline{--------------}

\begin{theorem} Let $\tilde{T} _ n(t)$ be the analog of $T_n(t)$ of Theorem 3 but now \underline{including upper limits}, the whole megillah.  Again integrating over all times, summing final states over $(\La)^n$, keeping one initial vertex fixed and summing the other vertices at $t=0$ over points lying in $(\La)^{n*}$.  Then 
\be \lim_{t\ra \infty} \tilde{T} _ n(t) = A_{n-1} \ee
\end{theorem}

\centerline{--------------}

\underline{On the Proof}

The computation of the right side of (42), arising as a solution of (40), was perhaps the most difficulty and tricky business I have ever been associated with.  Also I would find it extremely difficult to write a presentable proof.  Perhaps someone can come up with a reasonable proof.  (Skeptics may prefer to call Theorem 4 a conjecture, but it is certainly true.)  I content myself here with some points on the computation of $A_2$.

Equation (40), for $i=2$ becomes
\be	A_2 + a_2 + 3a_1.	\ee
The contributions of all contributing diagrams,when only the lower limit of the $t_2$ integration is kept, is $a_2 = -1$, the first term in (43).  This by Theorem 3.

Keeping the upper limit on the $t_2$ integration involves us with three cases.

1)  Case 1, associated to $t_1$ is $V_{z_1,z_2}$, associated to $t_2$ is $V_{z_1, z_3}$.

2)  Case 2, associated to $t_1$ is $V_{z_1,z_2}$, associated to $t_2$ is $V_{z_2, z_3}$.

3)  Case 3, associated to $t_1$ is $V_{z_2,z_3}$, associated to $t_2$ is $V_{z_1, z_2}$.

With this notation there is a sum over $z_2$ and $z_3$ with the restriction $z_2 \not= z_3 \not= z_1 \not= z_2$.  Here the contribution of $z_1, z_2, z_3$ does not equal the contribution of $z_1, z_3, z_2$.

Each of these three cases contributes a factor $a_1 = 1$ to equation (43).  Case 3 is the most interesting, and we will deal with this one case.

The contribution of Case 3 may be represented as
\be \sum_{x_1,x_2,x_3} \ \sum_{z_2,z_3} \int^t_0 \; dt_1 \; \int^t_{t_1} dt_2 \; K(x_1,x_2,x_3,z_1,z_2,z_3,t_1,t_2) \ee
$x_1,x_2,x_3$ lie in $\La^3$ and $z_1, z_2, z_3$ are restricted to $\La^{3*}$.  $x_1,x_2,x_3$ are positions of the vertices at $t=t$ and $z_1, z_2, z_3$, the positions at $t=0$.  Recall we are keeping only the upper limit in the integral over $t_2$, getting
\be \sum_{x_1,x_3} \ \sum_{z_2,z_3} \int^t_0 \; dt_1 \;  k(x_1,x_3,z_2,z_3,t_1) \left( e^{\Delta t}\right)_{z_1,x_1}  \ee
$k$ is the kernel of $a$ two-vertex diagram with a single interaction at $t=t_1$.  We rewrite this as 
\be  \sum_{x_1}  \left( e^{\Delta t} \right)_{z_1,x_1} \left( \sum_{x_3} \; \sum_{z_2,z_3} \int^t_0 \; dt_1 \;  k\left(x_1,x_3,z_2,z_3,t_1 \right) \right). \ee
We wish to compare this expression to
\be  \sum_{x_1}  \left( e^{\Delta t} \right)_{z_1,x_1} \left( \sum_{\bar x_1, x_3} \; \sum_{z_3} \int^t_0 \; dt_1 \;  k\left(\bar x_1,x_3,z_2,z_3,t_1 \right) \right).   \ee
\be \ \ \ \ \ \ \ \ \ \ \cong \ \ 1 \ \ \ \ \cdot \ \ \ \ 1.	\ee
by Theorem 3.  Using translation invariance of the kernel $k$, (46) and (47) differ by an error that goes to zero with $t$ (from the different restrictions on the $z$'s  in the two expressions).  In (47) we neglect the restriction that $z_2$ and $z_3$ may not equal $z_1$.

\bigskip

\centerline{--------------------}

\bigskip
We turn to the relation of Conjecture 1, equation (18)
\be   \lim_{t\ra \infty} \left( \sum_{\vec{x} \in A} f^e(\vec{x}, t)\right)^{1/N} = \left(C_N\right)^{1/N}.  \ee

We work in the limit $N$ large, and $t \ra \infty$ (before $N \ra \infty)$.  We thus want
\be \lim_{t \ra \infty} \left( \sum_{{\vec x} \in A} f^e(x,t) \right)^{1/N} = e  \ee
and eschew considering uniformity of $t$ limit with respect to $N$.  We view $f^e(x, t)$ expressed a sum of products of connected diagrams.  For $N$ large we expect this sum to be dominated by terms with some fixed number of connectivity patterns.  That is, in terms kept in the product there are $x_1N$ two-connected terms in the product, $x_2N$ three-connected terms in the product, $x_3N$ four-connected terms in the product, and so on.  The expression for $\displaystyle{\sum_{{\vec x} \in A}} f^e(x,t)$ in this limit is
\vfill\eject

\[	\prod_i \left( \frac{A_i \; i!}{N^i} \right)^{x_iN} \ \cdot \ \frac{N\;!}{(\sum_i(i+1)Nx_i)!(N - \sum (i+1)Nx_i)!} \ \cdot \]
\be \cdot \ \frac{(\sum(i+1)Nx_i)!}{\Pi(x_iN)!\;\Pi\left( (i+1)!\right)^{x_iN}} .  \ee
The first set of parentheses includes the contributions of the diagrams $t \ra \infty$ limit, from Theorem 4.  The next factor, a ratio of factorials, sums over which set of initial vertices are included in the set that are connected to other vertices.  The final ratio of factorials sums over the connectivities of the vertices (which vertices are connected with which vertices).

Maximizing (51) over the choice of $x_i$ one finds
\be	\lim_{t \ra \infty} \ \left(\sum_{{\vec x} \in A} \; f^e(x,t) \right)^{1/N} \ \cong e^q \ee
with
\be	q = -1 + \sum_{i=0} \ A_i \frac{p^{i+1}}{i+1} - \ell n \ p  \ee
where
\be		1 = \sum_{i=0} \ A_i p^{i+1} .	\ee
One wants $q=1$ but (53) and (54) do not yield $q=1$.  (It is not clear how to define a solution of (54) for $p$, but no reasonable definition works.)  At this point we have reached complete frustration!

Sometime after arriving at this impasse, we decided to consider a random walk not involving all the lattice vertices, but rather a fraction $\rho$ of the vertices, ``uniformly distributed".  This is achieved by ``integrating out" a fraction $(1-\rho)$ of the vertices in the probability function $f(x,t)$.  This is done before the extension to $f^e\;!$.  It is easy to make the corresponding changes in all the computations of this paper, a matter of a day or two.

Replacing (52), (53) and (54) one finds
\be	\lim_{t \ra \infty} \left(\Sigma \; f^e(x,t)\right)^{1/N} \cong e^{\tilde q}	\ee
with $f^e$ here depending on $\rho N$ vertices and with
\be  \tilde q = - 1 + \sum_{i=0} \ A_i \; p^{i+1} \ \frac{\rho^i}{i+1} - \ell n \; p	\ee
and
\be	1 = \sum_{i=0} \ A_i \; p^{i+1} \rho^i .  \ee
Where before one wanted $q=1$, here we desire
\be  \tilde q = 1 + \frac{1 - \rho}{\rho} \ \ell n (1 - \rho).	\ee
The miracle that happens is as follows.  For $\rho < 1/2$, equation (57) is satisfied with
\be	p = 1 - \rho   \ee
and substituting this expression for $p$ into (56) one finds (56) and (58) yield the same formal expansion in powers of $\rho$, valid for $\rho < 1/2$.  It is interesting to note that (56), (57) and (58) determines both (59) and the $A_i$ (expanding $p$ and $\tilde q$ in powers of $\rho$).

If we set
\be    f(\rho\, p) \equiv \sum_{i=1} \ A_i \; p^i \rho^i  \ee
we can solve, from (57) and (59) 
\be	p + pf(\rho\, p) = 1 \ee
\be p = 1 - \rho	\ee
to get
\be	f(z) = \frac{1 - \sqrt{1 - 4z}}{1 + \sqrt{1 - 4z}}	\ee
and see the singularity at $z = 1/4$, or $\rho = \frac 1 2$.

So we are led to believe that Equation (6) may still be true in some suitable sense, but that a perturbation expansion development as undertaken in this paper is not promising.  Presumably as $\rho$ increases to value $1/2$ one must consider diagrams of arbitrarily high connectivity.

Many interesting questions suggest themselves, of which we choose two.  For $\rho < 1/2$, control the perturbation expansion, and obtain uniformity in $N$ of the $t \ra \infty$ limits.  Find some way of treating $\rho > 1/2$.  The first question is likely a problem about which to develop several Ph.D. theses.  The second still requires some further ideas to gauge its difficulty.

\vfill\eject

\centerline{\underline{References}}

\begin{itemize}

\item[[1]] Robert T. Powers, ``Heisenberg Model and a Random Walk on the Permutation Group", {\it Lett. in Math. Phys.} {\bf 1}, 125-130 (1976).
\item[[2]]  P. Federbush, ``For the Quantum Heisenberg Ferromagnet, Tao to the Proof of a Phase Transition", math-ph/0202044.
\item[[3]]  David C. Brydges, ``A Short Course in Cluster Expansions, Phenomenes critiques, systems aleatoires, theories de gauge, Part I, II" (Les Houches, 1984), 129-183, North Holland, Amsterdam, 1986.

\end{itemize}

\end{document}